# Reconfiguring and ramping-up ventilator production in the face of COVID-19:

# Can robots help?

**Ali Ahmad Malik, Tariq Masood and Rehana Kousar**

**University of Southern Denmark, University of Cambridge and Cambridge Biomakespace**

**Pre-print**

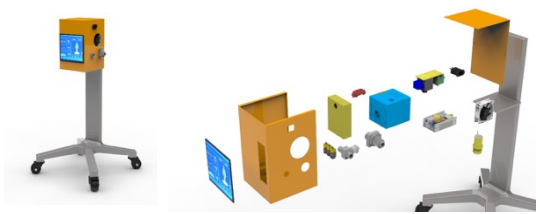
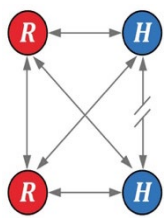
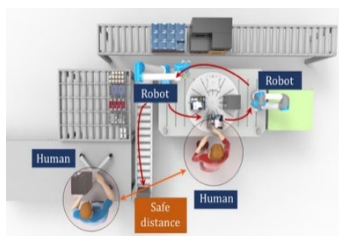
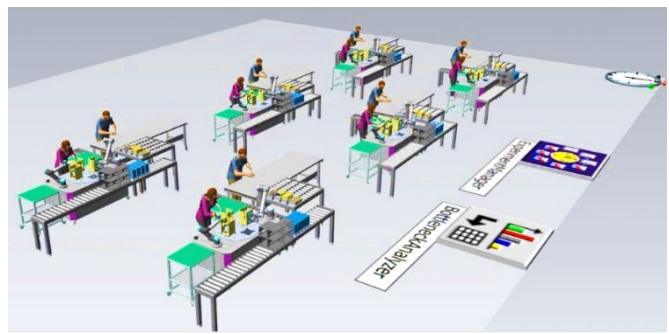

## Highlights

- COVID-19 pandemic has limited almost all aspects of life with social distancing measures in place
- Among others, medical supplies particularly ventilators are life savers but in huge shortage
- However, current ventilator production capacity cannot cope with the huge demand in short time
- Reconfiguring and ramping-up ventilator production in non-traditional ways is the great need of the hour
- 11-point guidance on the use of collaborative robots including social distancing measures is provided

---

# Reconfiguring and ramping-up ventilator production in the face of COVID-19: Can robots help?

Ali Ahmad Malik[1*], Tariq Masood[2,3,4**] and Rehana Kousar[5]

[1]Mads Clausen Institute, University of Southern Denmark, Sønderborg, Denmark

[2]Institute for Manufacturing, Department of Engineering, University of Cambridge, 17 Charles Babbage Road, Cambridge CB3 0FS, UK

[3]Engineering Design Centre, Department of Engineering, University of Cambridge, Trumpington Street, Cambridge, CB2 1PZ, UK

[4]Cambridge Global Challenges Strategic Research Initiative, University of Cambridge, JJ Thompson Road, Cambridge CB3 0FS, UK

[5]Biomakespace, Cambridge University Biomedical Innovation Hub, Clifford Allbutt Building, Biomedical Campus, Hills Road, Cambridge, CB2 0AH, UK

*E: alimalik@mci.sdu.dk; **E: tm487@cam.ac.uk

***Abstract:***

*As the COVID-19 pandemic expands, the shortening of medical equipment is swelling. A key piece of equipment getting far-out attention has been ventilators. The difference between supply and demand is substantial to be handled with normal production techniques,* especially under social distancing measures in place*. The study explores the rationale of human-robot teams to ramp up production using advantages of both the ease of integration and maintaining social distancing. The paper presents a model for faster integration of collaborative robots and design guidelines for workstation. The scenarios are evaluated for an open source ventilator through continuous human-robot simulation and amplification of results in a discrete event simulation.*

## 1. Introduction

The world today is facing COVID-19 pandemic that, since its first case appeared in Wuhan, China in Dec 2019 until 18th of June 2020, has already affected 8,0 million people (confirmed cases) with 441,290 confirmed deaths in 213 countries, areas and territories (see Figure 1) [1]. The pandemic is further expanding and has already led to huge disruption in almost every aspect of daily life. One of the most impactful shortage is of medical supplies. Technically, the most complex product among the shortage-facing equipment is medical ventilator that is used for artificial breathing when a patient, in severe cases of COVID-19, becomes unable to breath naturally [2]. However, one of the biggest global challenges is that the number of ventilators available in hospitals are considerably less than required [3] for effectively dealing with the pandemic due to a widespread shortage of medical ventilators and protective gears [4].

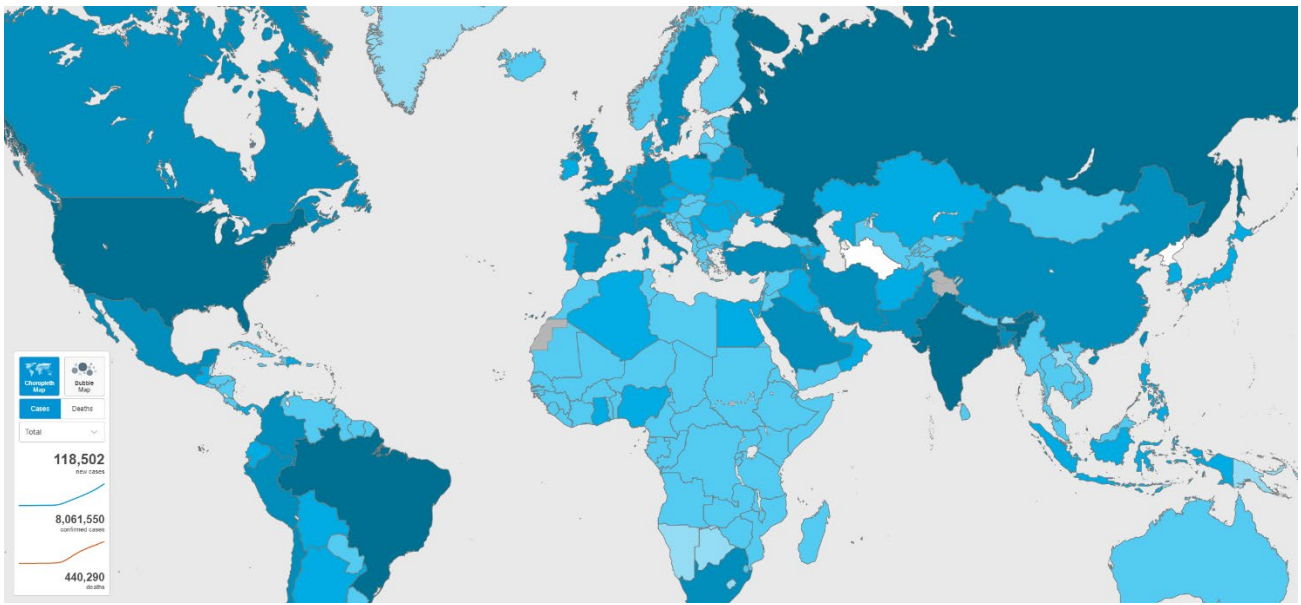

*Figure 1: World map of confirmed COVID-19 cases as of 18th June 2020, 01:00 BST (WHO 2020).*

Due to the unprecedented circumstances and ventilator shortages, the governments have been putting manufacturers into war footing to produce additional ventilators to meet the demand [5]. The companies have been aiming to boost their productions, e.g. Hamilton and Getinge, comparing to their current production of 15,000 and 10,000 units, are aiming to produce 21,000 and 16,000 ventilators respectively in the 2020 [6]. Non-medical device manufactures, on the other hand, are taking rapid measures to repurpose their production lines to tackle the unprecedented demand as a social responsibility or on calls of the governments, e.g. Ford and General Electric (GE) have aimed to produce 50,000 machines in 100 days [7]. Corporate giants, Siemens AG and Airbus SE, have also responded positively to address the demand of ventilators [4]. Dyson has designed a new ventilator in 10 days, and has plans to produce 15,000 ventilators [8].

But the momentous demand is difficult to achieve as the manufacturing systems of today, aimed for mass production, are designed with automation solutions not flexible enough for large scale reconfigurations [9][10]. There is a lack of general-purpose automation solutions [11]. Furthermore, there is a high human involvement in many processes such as assembly [12, 13] while assembly is by far the dominant activity in ventilator production [13]. It upholds another challenge in the time of a pandemic (particularly in the case of COVID-19) that adhering to **quarantine** rules require minimum people to go out to work; and (those who do) have to maintain **social distancing** guidelines [5]. Automation (e.g. robotics) is an effective way to replace human effort, relieving humans from physical tasks and boost production volume [14]. However, conventional automation is inflexible making it very difficult or almost impossible to repurpose the production lines on a very short notice, and often not able to coexist with humans.

A newer hardware in the class of industrial robotics is **collaborative robot or cobot** [15]. Cobots are safer for humans, easier to integrate and are suitable for automation of physical tasks while coexisting with humans [16][17]. Besides high potential of cobotic automation and cases available today of using cobots for single repetitive tasks (e.g. machine tending), their application in assembly is limited [18]. The aim of the research presented in this article is to explore the role that collaborative robots can play in reconfigurable automation for assembly of medical ventilators. A model for integration of cobots, recommendations for cobots based workspace design and recommendations for faster integration of cobots are suggested. Simulation results for cobot based production of an open source ventilator are presented and discussed before concluding this research.

## 2. THEORETICAL BACKGROUND
### 2.1. The COVID-19 pandemic

The defining global health crisis of our time is the novel coronavirus COVID-19, and undoubtedly the greatest challenge humans have seen since World War II [19]. "Coronaviruses (CoV) are a large family of viruses that cause illness ranging from the common cold to more severe diseases. A novel coronavirus (nCoV) is a new strain that has not been previously identified in humans" [1]. The WHO Director-General declared the novel coronavirus (2019-nCoV) outbreak a public health emergency of international concern on 30 January 2020 [1]. The virus is spreading like a wave in every continent and country except Antarctica. It took over three months for the novel coronavirus to affect the first 10,000 people while it took only 12 days to reach the next 100,000 [20]. The number of deaths as of now are 441,290 concluding a mortality rate of 5.5% [1]. Consequently, the disease was given the status of a *pandemic* on March 11, 2020 by the World Health Organization (WHO) [21].

The novel coronavirus, after getting into a human body, attacks the lower respiratory system and the victim may experience fever, dry and persistent cough, myalgia and shortness of breath [21]. While the life-threatening situation may result in the form of pneumonia and acute respiratory distress syndrome (ARDS). The WHO estimates that some 80% of the affected people recover without requiring any hospitalization [22] however, serious breathing difficulties are observed in one in every six COVID-19 victims [23]. In these severe cases the virus damages the lungs, in response the immune system of the body reacts by expanding the blood vessels and causing the fluid to enter into the lungs. It makes it difficult to breath, as a result of which, the oxygen level drops [22], and can lead to a fatality. The seriously ill people need to be treated in Intensive Care Units (ICUs), where medical ventilators are required to provide artificial breathing support.

### 2.2. Medical ventilators

Machine ventilation or artificial breathing is facilitated to a person experiencing difficulty or loss of all the ability to breath naturally [13]. A ventilator enables artificial breathing, through mechanical aids and oxygen under a defined period of pressure and volume (i.e. ventilation pattern) [13]. A ventilator is a device that gets oxygen into the lungs and removes carbon dioxide from the body (Figure 2) and is considered a common feature of an ICU [24].

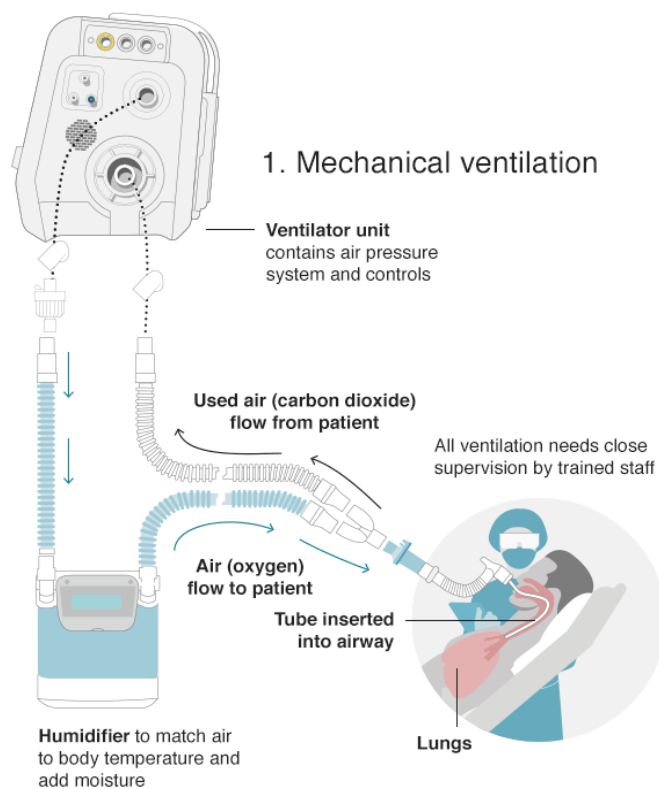

*Figure 2: A schematic of machine ventilation to a patient [22].*

The history of ventilatory assistance can be traced back to Biblical times [24] but the modern day ventilating machines dates back to 1940s [25]. Since then, the ventilators have evolved from totally machine-triggered volume-ventilating devices to microprocessor-controlled smart respiratory systems [24]. The journey is still continued to make them smarter, portable, less damaging to the lungs and data-connected to other equipment of an ICU.

### 2.3. Shortage of medical ventilators

Ventilating machines are a common feature of acute care beds in hospitals [26]. The number of acute care beds available in a hospital is considered an important indicator of health service quality [27]. The EU average of acute care beds available in hospitals is 11.5 per 100,000 heads of population (Figure 3), while for USA it is 28/100,000; as of 2010 [26]. As a matter of course, not all ICUs are equipped with ventilators. In USA, at the start of COVID-19 pandemic, 62,188 ventilating machines were available while the projected need was 960,000 [28]. Similarly, the hospitals in the U.K. were equipped with 8,000 ventilators while 30,000 ventilators were estimated to be needed at the peak of pandemic [23].

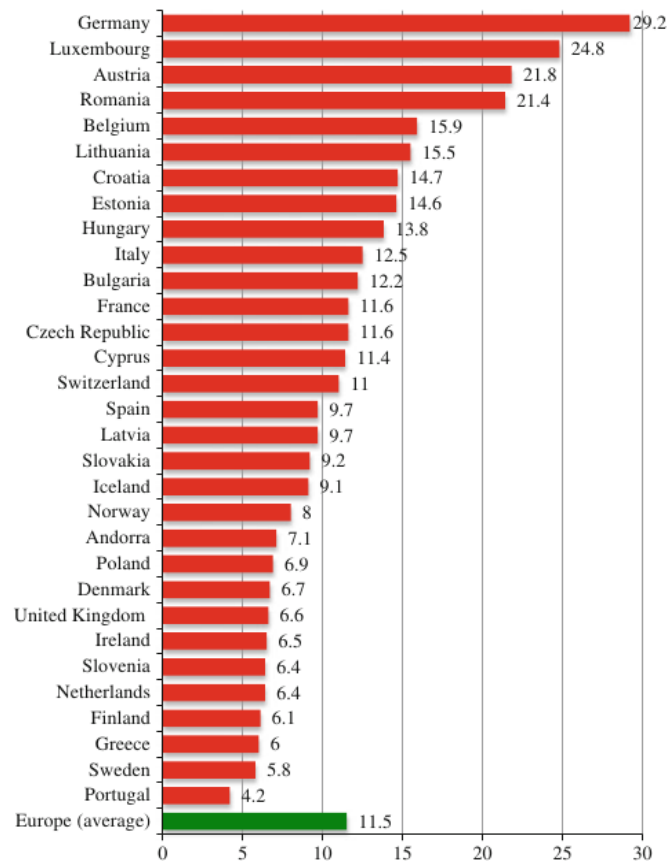

*Figure 3: Number of acute care beds available in Europe per 100,000 capita of population* [26].

The available ventilating machines in low- and middle-income countries are significantly lower, where the pandemic is just beginning to outsize except China where it appears to be coming under control now. The three most populated countries in South Asia i.e. India, Pakistan and Bangladesh, comprising nearly one quarter of the global population, have 3.7, 2.5, and 3.6 critical care beds respectively for every 100,000 population [29]. The situation is even appalling in underdeveloped countries, such as Mali, that has only 56 ventilators for a 19 million population [23]. To conclude, as of now, a big reason for mortality among COVID-19 patients has been described as the unavailability of machine ventilators.

## 2.4. Reconfiguring factories at the peak of a crunch

Gearing up the response efforts, the governments across the world are forming strategies together with the manufacturers to reconfigure their factories and to ramp up the production in order to meet the anticipated demand of ventilators. Manufacturing reconfiguration refers to quickly adapt to new tasks and requirements to increase volume or to produce a different product [30]. It can be the ability to add machines to existing operational system very quickly to respond rapidly and cost effectively to unexpected demand surge [31][32]. Koren [33] defines responsiveness of manufacturing systems as to rapidly and cost effectively producing new products on existing systems if facing market changes, product changes or if (a part of) a system fails.

The manufacturing systems of today are designed for mass production with minimum product variability at competitive cost [30]. Significant research has been produced in the past decades to build reconfigurable or responsive manufacturing systems [34][34, 35][36] but their realization in industry is undoubtedly limited. As a result, conventional manufacturing systems, even when exhibiting a degree of flexibility, are often not easy to up-scale or down-scale to react if the demand fluctuates [33].

The complications to realize responsive manufacturing systems have been the high initial cost and increase in systems' complexity. There is a lack of general-purpose automation solutions that can be integrated and reconfigured conveniently and quickly to boost automation and increase production volume. Correspondingly, in the time of current pandemic, only the leading companies equipped with gigantic technical and human resources are (hesitantly) aiming to address the need of fluctuating demands, while they face challenges of supply chain disruptions, large number of staff working from their homes, and social distancing measures globally but particularly where they work in factories.

Automation increases production volume, improves product quality and relieves humans from repetitive tasks [37]. Industrial automation of physical tasks is greatly driven by robots in today's manufacturing [38]. However, robots remain too inflexible [36] and are impractical when need to coexist with humans [39]. It takes weeks to integrate a robot in a production system; with little to no possibility of any modifications without considerable efforts.

Furthermore, a substantial production ramp-up amid COVID-19 pandemic is needed when humans are asked to maintain social distancing or preferably stay at home. Only, the least possible staffing, is allowed at production floors or assembly lines while maintaining a distance of minimum two meters from each other (as currently advised by some governments) aimed at minimizing the spread of COVID-19.

## 3. COBOTS AS COWORKERS

In recent years there has been a notable interest in deploying robots as coworkers, referring to them as collaborative robots (aka cobots) [40][41][42]. Cobots have emerged as a newer hardware in the class of industrial robotics (Figure 4). They don't require closed-off areas for their operation, instead they can cooperate or collaborate with humans and share their workload to perform tasks requiring a combination of their best and complementing competencies [17][18]. Furthermore, they are easier to program (or repurpose) and are portable [43]. Cobots have been successfully integrated into several industrial applications in the past years from simple to complex tasks [44] but their full potential in labor intensive tasks, such as in assembly, has remained untapped.

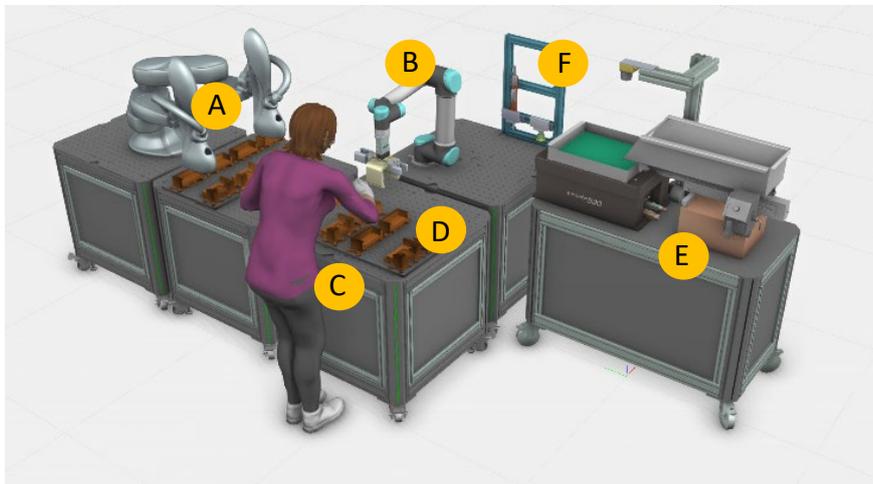

*Figure 4: Schematic of human-robot collaboration for assembly work.*

The key idea with cobots is the partial automation of some of the tasks (that doesn't necessarily require human skills) of a process in a way that cobots tend to assist or empower the fellow human [45]. Cobots are aimed at increasing the degree of automation in manual work, thus opening new applications or domains for automation. Since assembly is among the most labour-intensive elements of a manufacturing value chain, the automation of a fraction of it can potentially has huge benefits [46].

## 4. RAMPING-UP VENTILATOR PROUCTION WITH HUMAN-ROBOT ASSEMBLY SYSTEMS

A responsive and evolvable HRC production system for medical ventilators can be designed exhibiting a modest initial production capacity that has the ability to add additional production capacity and functionality as the demand increases (Figure 5). The two capabilities needed for a reconfigurable production system are: (i) the ability to change the functionality, and (ii) the capacity, while the critical design issues are related to the layout of its elements or stations and the assignment of tasks to these stations [33].

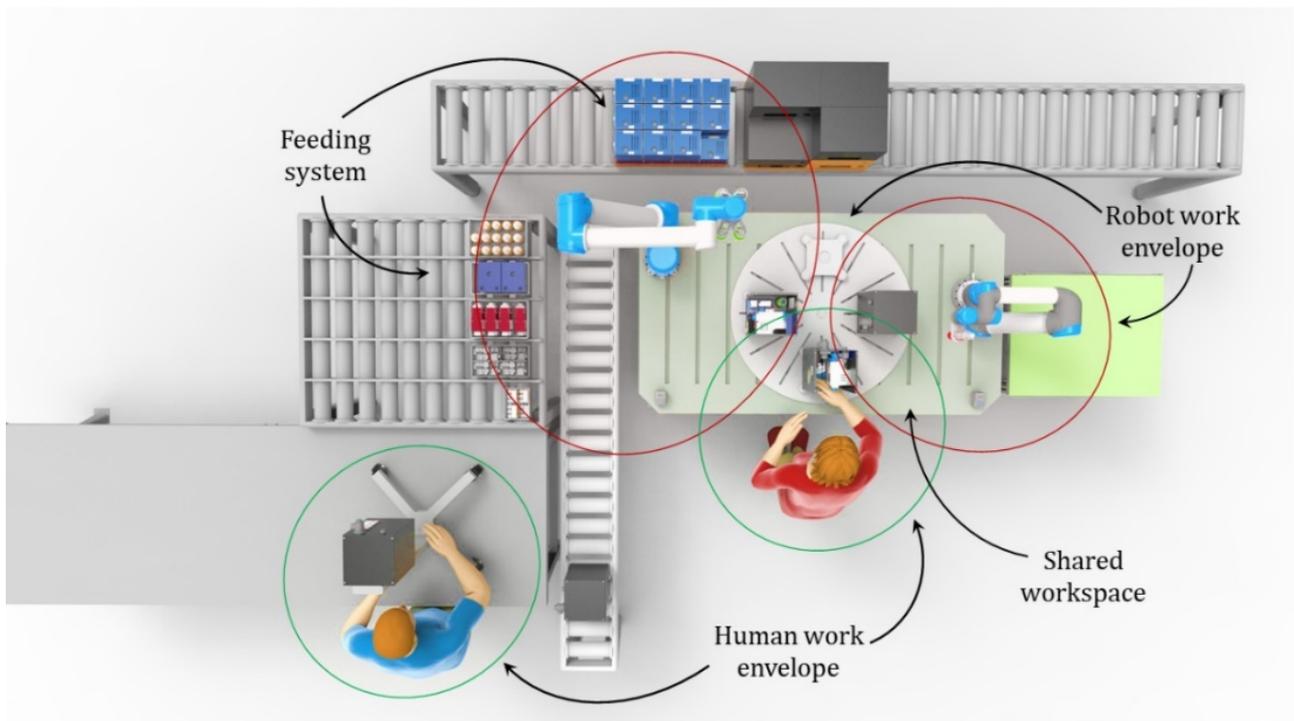

*Figure 5: Schematic of human-robot assembly system producing medical ventilators.*

The production of ventilators (similar to many other products) is largely an assembly activity. Assembly is conventionally a manual activity due to its complexity and variety of tasks that are deemed suitable for humans [47]. A large proportion of these tasks are repetitive and easier to automate by a robot; but due to the complexity of the remnant tasks and the safety challenges of coexistence of humans and robots, the whole process often remains highly manual [38].

Depending upon the design of ventilators, its assembly may have several pick-and-place tasks that can be performed by a robot while coexisting with humans who are taking care of the tasks that require human skills and dexterity.

The process of designing an HRC assembly system starts from decomposing the assembly process into tasks and evaluating each task for its automation potential (see Figure 6). Thereby the tasks with a high automation potential for a cobot are separated from the rest of tasks. To balance the assembly process between human and robot, it is required to know the cycle time for the manual as well as robotic tasks. This can be achieved through computer or physical simulation. Finally, a task allocation can be carried out considering assembly precedence and cycle time.

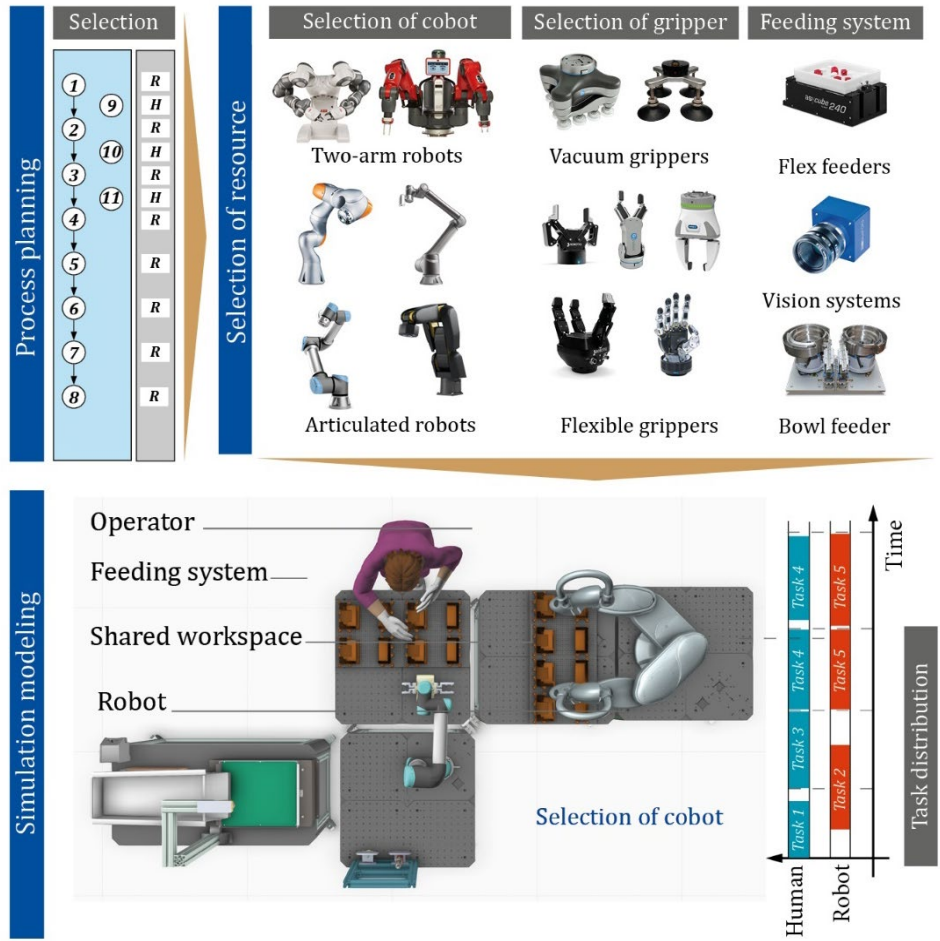

*Figure 6: The steps in designing an HRC assembly system.*

According to the assembly process, the hardware and human resources are selected, which includes the selection of both the active resources (e.g. human operator, the robot manipulator and its tooling etc.) and passive resources ( e.g. tables and fixtures). The selection of the robot and tooling may have various views ranging from technical, economic and social [43]. The technical aspects are the payload, the reachability, the capability of handling the components, the number of resources required for a task, etc. [48]. The right robot manipulator is context specific as different robot designs along with its accessories have different advantages [49].

The layout planning involves the selected hardware to be arranged in the workspace. Layout planning is used to organize the resources in the given space considering the interaction between the active and passive resources. The attributes that form a decision-making arena for the comparison between layout alternatives are travel distance of operator and material, safety, aesthetics, operator acceptance, ergonomics, throughput etc. [50]. Simulation tools with digital human and robotic models can be used to design and evaluate workplace layouts for HRC [48].

Additional planning may involve cobot programming and safety system design. Once an initial sketch is ready, a preliminary design of the HRC system can be conceived. A set of generic **guidelines** are presented for design of an HRC workspace for ventilator production focussed at ramping up production.

1. **Keep it modular:** Since the cobot stations are desired to be flexible, modular fixturing tables can be used to enable easy (dis)mounting of fixtures and other hardware.
2. **Keep it small:** Miniaturize the cell as much as possible. A large workspace will require the cobot as well as the operator to make unnecessary movements.
3. **Logical flow:** Determine a logical and sequential flow of the product. A human-cobot collaborative cell has a high potential to enable one-piece flow, and minimization of any work-in process.
4. **Ergonomics:** Ensure that the workstation is ergonomic friendly, and the operator doesn't need to bend the body to reach its target locations.
5. **Visibility of the cobot movements:** Place the cobot in the eye-view of the operator.
6. **Emergency stop:** An emergency stop should always be present in the arm reach of the operator.
7. **Human interaction with cobot:** For every task execution by the cobot, an enabling device must be available to the operator.
8. **Minimize collisions:** Determine the cobot workspace and operator's workspace and minimize their overlapping.
9. **Maintain social distance:** Keep recommended social distance between human operators on the HRC assembly stations (e.g. 2-metre as currently recommended by some governments)
10. **Maintain a digital twin:** Validate the cell and its performance using simulations, and retain it as a virtual platform for validation of future modifications
11. **Simpler design:** Keep it simple, easy to develop and reconfigure

Cobots can be a great choice for **repurposing the existing production systems for making ventilators** in terms of:

- Reducing direct man-hours needed to produce the ventilators,
- Reducing total time for the required production volume,

- Safer colleagues of humans to maintain social distancing,
- Saving the investment as the cobot can be easily repurposed for another application once the pandemic is over, and
- Standard robot program templates can be developed for faster integration.

## 5. SIMULATING HRC HUMAN-ROBOT ASSEMBLY SYSTEM FOR VENTILATOR PRODUCTION

The presented scope of cobots for ramping-up of ventilator production is examined through computer simulations, which helped in:

1. Performing ventilator assembly tasks by the cobots,
2. Producing detailed design and layout planning of the workspace, and
3. Stochastic analysis of the HRC production system.

Two different simulation environments were used to evaluate cobots for production ramp up for ventilators. Firstly, a dynamic time-lapsed simulation was performed based on continuous simulation (CS) techniques. The robotic and human tasks were executed in this simulation for detailed process visualization, and cycle time evaluations. Secondly, the assembly process was balanced using this simulation avoiding any idle times.

The results were then exported into discrete event simulation (DES). The rationale behind using two types of simulations is proposed by [51] as CS accounts for time-dependent dynamic behaviour of the system while DES includes stochastic effects. By using the two simulations the advantages are combined while minimizing the shortcomings of each.

### 5.1. Mechanical ventilator and assembly process description

The case used to demonstrate the usability of cobots forming a human-robot assembly system to produce medical ventilators is shown in Figure 7. The ventilator, in the use case, consists of 15 unique parts and each part defines a unique assembly task. There can be additional testing and orientation tasks involved. The case is based on an open source design of ventilator [52] however; some design modifications were made to make it convenient for an automated assembly.

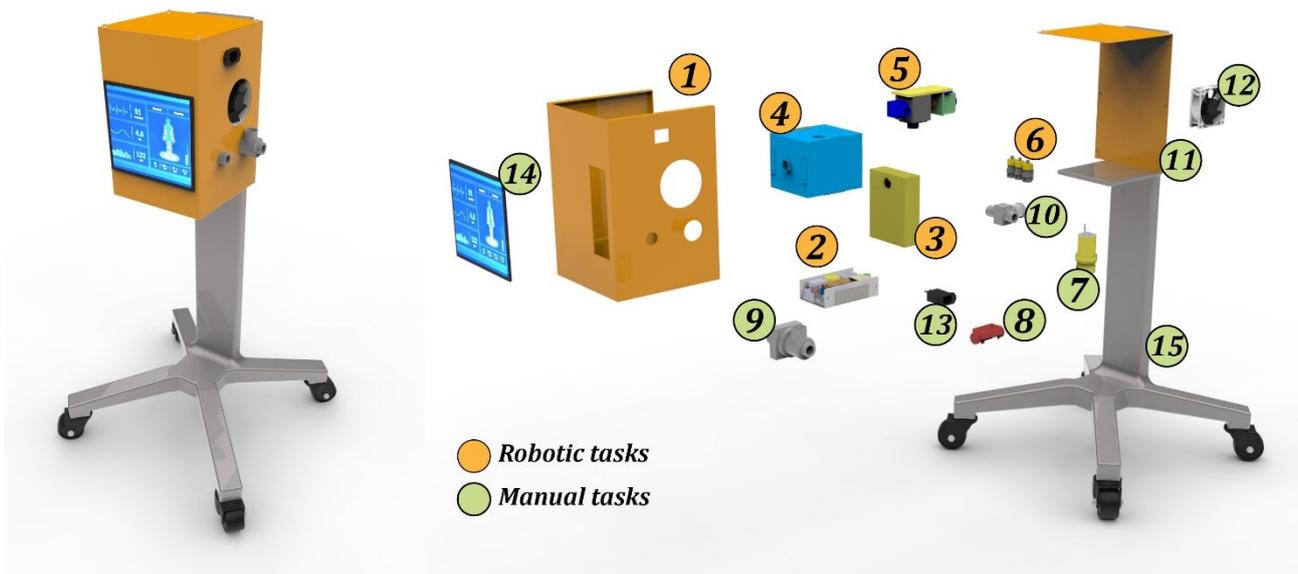

*Figure 7: Mechanical ventilator parts.*

The assembly station (see Figure 5) is comprised of two robot-manipulators and two humans. This robot manipulator used in this study is Universal Robot UR-10e. The robot has six degrees of freedom, a payload capacity of 10 kg and a reach of 1,000 mm. A vacuum gripper by EPIC is used with the robot. The second robot is a Universal Robot UR-5e. The robot has six degrees of freedom, a payload capacity of 5 kg and a reach of 850 mm. The robot is equipped with a screwdriver at its tool post to perform the screwing tasks.

### 5.2. Work distribution and process balancing

The complete assembly process is decomposed into tasks, where each task is evaluated for its ease of cobotic automation. The evaluation of tasks for HRC automation is different from conventional robotic automation as safety implications also need to be considered. Different techniques have been described in the literature to evaluate product complexity for cobot automation [53][54][55][56][57].

The most important factors constituting complexity of a task for automation include the shape of the component, feeding of parts to the assembly station, the alignment needed between two mating components, and the risk of injury for the humans [18]. A simplified method for task distribution is shown in Table 1. If a task has no potential challenge in any of the attributes (i.e. handling, feeding, mounting, joining and safety) for performance by the robot; it denotes that the task can be performed by the cobot. The remaining tasks can be assigned to the human.

*Table 1: Evaluation of ventilator assembly tasks for assignment to cobot.*

| Task name | Assembly attributes | Rating 1/0 (1= robot can do; 0= robot can't do) | Resource |
|---|---|---|---|
| **Task 1** | Physical shape (S) | If robot can handle the part | If (S && F && M && J && Sf == 1, "Robot"; else "Human") |
| | Part feeding (F) | If parts have a structured presentation to robot | |
| | Mounting (M) | If no precise adjustment is needed | |
| | Joining (J) | If joining task is easy for the robot | |
| | Safety (Sf) | No safety risk if performed by the robot | |

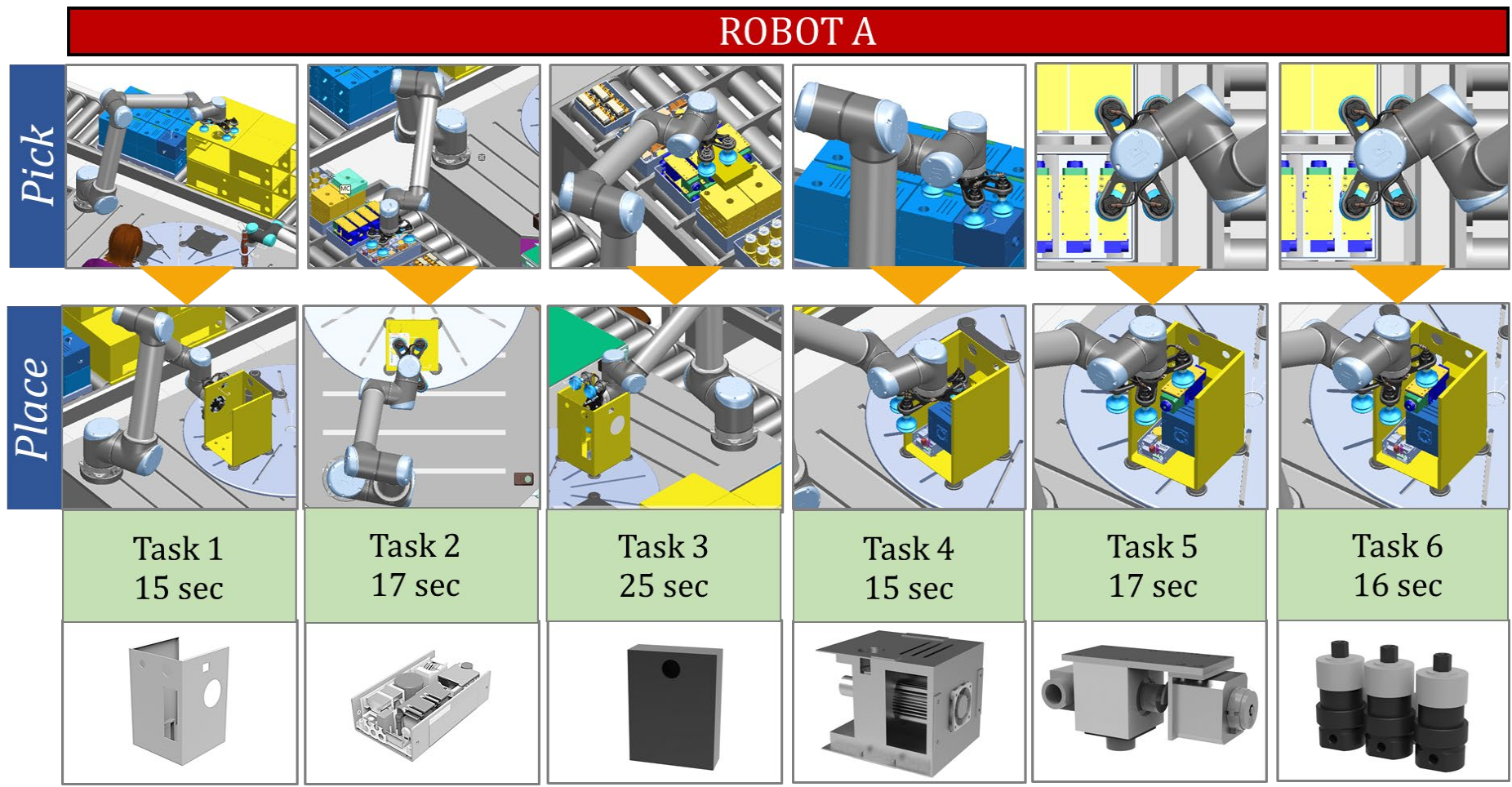

*Figure 8: Ventilator assembly tasks assigned to the cobot.*

The assembly process follows a sequence that is referred to as assembly precedence [58]. To distribute the tasks between human and robot, task complexity, as well as the precedence need to be considered. In the ideal situation, the tasks assigned to the cobot must take same amount of time as the predecessor or successor human tasks to have maximum utilization of resources.

The analysis and simulation show that nearly half of the assembly tasks can be performed by the cobot. Six of the total fifteen assembly tasks were selected to be assigned to the cobots (Figure 8). Primarily, these tasks included picking and placing of components (Task 1 to 6) and mounting of screws. The remaining nine tasks were kept as manual.

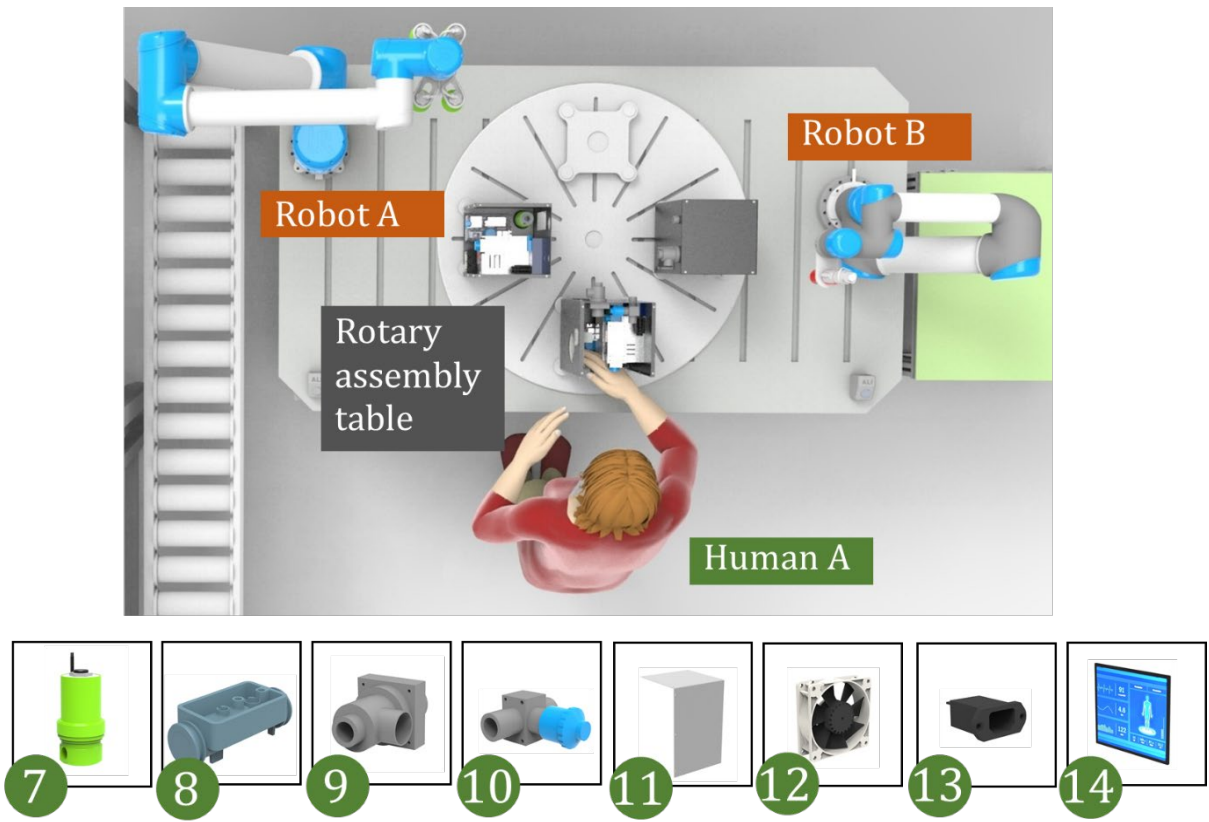

*Figure 9: Ventilator assembly tasks assigned to the human.*

Robot-A initiates the process and completes tasks 1 to 6 in 106 sec. Then the assembly table rotates, and the work-in-process is moved to the human. Human-A performs subsequent tasks 7 to 14, taking 140 seconds. After this, the assembly station rotates again, and the Robot-B mounts all the screws. The assembly table rotates and reaches to Robot-A that picks and places the assembled product on the delivery conveyer and starts working on the next product. Human-B picks the sub-assembly from the conveyer and conducts tests and then attaches the stand to it. It takes 406 seconds to complete one unit of ventilator (Figure 10).

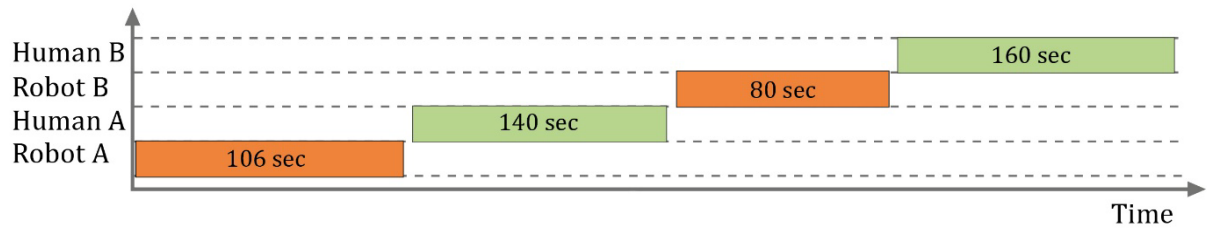

*Figure 10: Cycle time to produce one unit of ventilator.*

There are two operators being used at each station. One of the operators (Operator-A) directly collaborate with the robots to perform the assembly tasks (Figure 9). While the operator-B performs the final testing, and assembles the stand with the product, and places the final assembly on a trolley.

### 5.3. Production ramp-up for ventilators

The HRC system is then amplified with multiple workstations (Figure 11) incorporating stochastic variables to quantify the sensitivity of the results under production variables (Table 2). The cycle time (CT) to produce one unit of ventilator is derived from the kinematic simulation in section 5.2. Process time (PT) includes waiting/ idle time. Setup time (ST) is the time required to prepare the station before it starts its operation. The time to feed the parts into assembly stations is taken as setup time, and the availability (AV) of each resource is taken as 100%.

*Table 2: Variables concerning ventilator assembly processes and operators.*

|         | CT (sec) | PT (sec) | ST (sec) | AV   |
|---------|----------|----------|----------|------|
| Robot-A | 106      | 114      | 1800     | 100% |
| Human-A | 140      | 152      | 1800     | 100% |
| Robot-B | 80       | 92       | 1800     | 100% |
| Human-B | 80       | 160      | 1800     | 100% |
| **TOTAL** | 406    | 518      | 7200     |      |

Available working time per shift is 7 h and 30 min = 450 min = 27,000 sec. The results are simulated for a demand of 5,000 ventilators to be produced in 15 days, which sets a target of minimum 334 ventilators per day. The takt time is calculated by dividing the available working time by the number of ventilators needed (i.e. 80 sec).

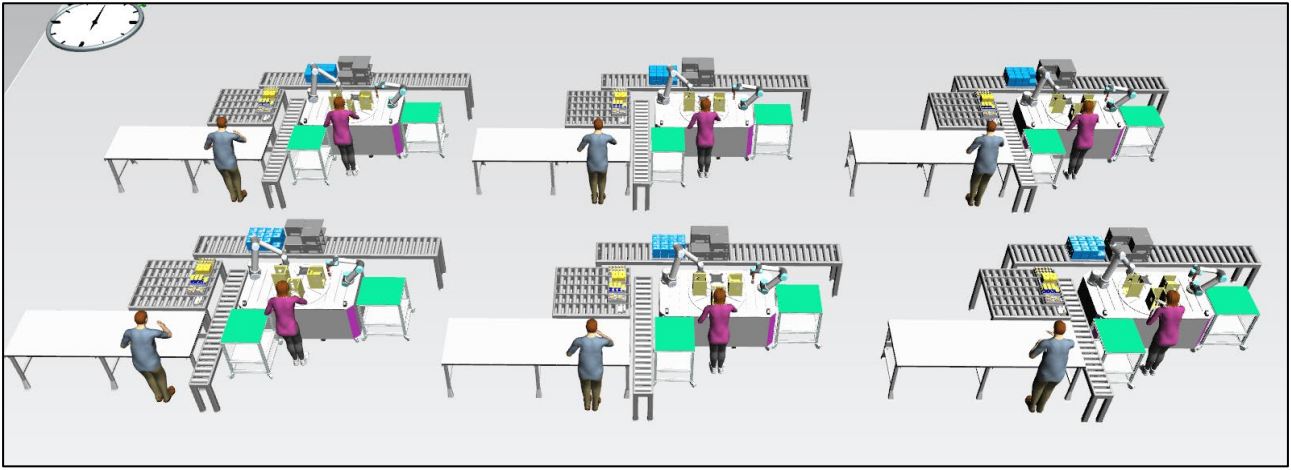

*Figure 11: Ramping-up ventilator production - DES simulation with stochastic variables.*

From the DES results the average throughput is 85 ventilators per day from one HRC station. It will require 4 HRC assembly stations to produce 340 units of ventilators in a day. It shows that four HRC assembly stations can produce 5,100 ventilators in 15 days.

## 6. DISCUSSION

As the pandemic of COVID-19 progresses rapidly, several questions of socio-technological life need to be urgently addressed. Considering that every crisis brings a sense of responsibility, it is hoped that the COVID-19 pandemic will make humankind recognize the need of sustainability and the value of joint actions to fight global challenges [59]. In terms of meeting demand of medical supplies, shortage of medical ventilators has posed one of the most pressing challenges. In the given case, the use of cobots has demonstrated help in:

- Automating nearly half of the processes required for ventilator assembly,
- Decreasing man-hours required to produce the same or more production volume of ventilators,
- Maintaining social distance between humans as required at times of pandemics, and
- Reconfiguration of the cobots, which may later be used for some other tasks easily.

### 6.1. (Re) Design of the workspace

A carefully (re) designed assembly workstation results in lower cost, shorter cycle time, ease of reconfiguration and safety. A set of guidelines for a human-robot collaborative assembly workstation is formulated. These guidelines are based on the design guidelines for lean workstations [60], safety specifications of ISO15066, and authors' experiences. Guideline 1-5 are focused on human-cobot interaction while guidelines 6-8 and 10-11 are identical to the lean design of the assembly workstations. Guideline 9 related to social distancing is unique to the current situation of a pandemic and would be applicable in such unprecedented circumstances.

### 6.2. Social distancing

The COVID-19, accustomed to history of pandemics [61][62], has brought strict social distancing measures. In addition to their international border closures and suspension of air traffic, the governments are using national guards to restrict the public movement [5]. The governments around the globe have asked the public to limit their social contacts as much as possible. If someone gets symptoms of COVID-19, the whole household is obligated to self-quarantine for 14 days [63]. Furthermore, people in high risk category (i.e. older than 70 years, pregnant women and people with history of severe diseases e.g. severe asthma, lung diseases, and hypertension etc.)  are required to self-isolate [64]. With a higher compliance to the strategy, the pandemic can likely buy more time by slowing down the spread and finally be averted with other measures in place, particularly vaccines [61].

Social distancing is targeted for the time until a strain-specific vaccine is developed and distributed which may take several months or even years. The implementation of social distancing strategies for such a  long time is very challenging. It brings halt to economic activity, mass unemployment, and shortage of goods. Factories, besides producers of medical supplies, are asked to remain operational with minimum possible staffing.

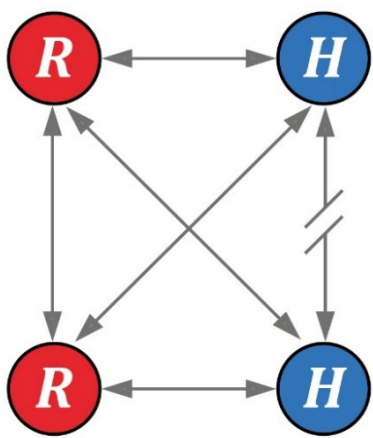 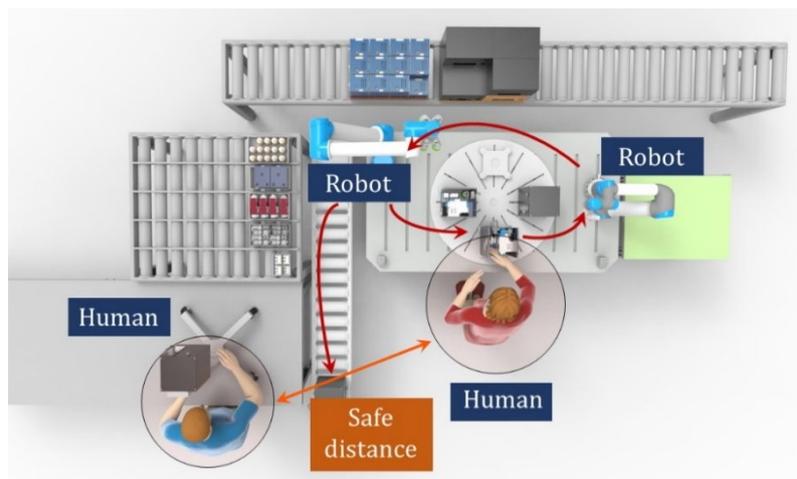

*Figure 13: Social distancing measures adapted with human to robot interaction only and maintaining a distance of 2m  between the two operators.*

Equivalently, maintaining social distancing strategies in factory areas is also challenging. Most manufacturing companies are operational while maintaining a distance between operators during and after work. In this scenario, cobots, machine colleagues of humans, are inherently able to support the social distancing strategies. It is suggested that the work tasks are designed in a way that the direct interaction between humans is made limited, rather interaction is made between robot and the human where possible (Figure 13).

The arrows in the given figure show possible engagements between human(s) and robot(s). The engagement between human to human is broken down as a social distancing measure. Even with this model of HRC engagement, transfer of a pathogen is possible from surface of a work-in-process that is routed from an infected human to a robot and then to another human. In order to de-risk this, wearing hand gloves must be ensured for all human operators.

### 6.3. Enablers for reconfiguring factories through cobots

Manufacturers hold a history of facing harsh market fluctuations in the face of socioeconomic crunches often requiring them to upscale or downscale their capacity or to make products they are not experienced with. The pressure to adapt to the demand changes gets intensive for certain products and for their regular (and closely related) manufacturers. The research presented in this article has proposed HRC as a potential solution for flexible and reconfigurable automation for tasks such as assembly.

#### 6.3.1. Modularization

A modular production system constitutes a universal architecture both in the mechanical design of its elements (modules) as well as in the information flow. The universal architecture can enable a quick and economic inclusion or exclusion of the elements of a production system for capacity or capability adjustment [65]. Modularization maybe achieved both in the product being produced and in its manufacturing system [66] [67]. A reconfigurable HRC assembly cell may address demand fluctuations if the modularization is integrated in its hardware (robots, robot-tables, conveyors, fixtures) and software (robot programs) [65]. If the demand increases, additional modules may be added to the existing system with little effort. The approach is analogous to the LEGO construction bricks (Figure 14).

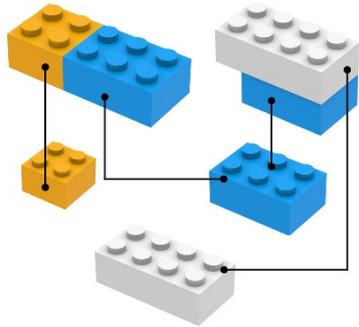 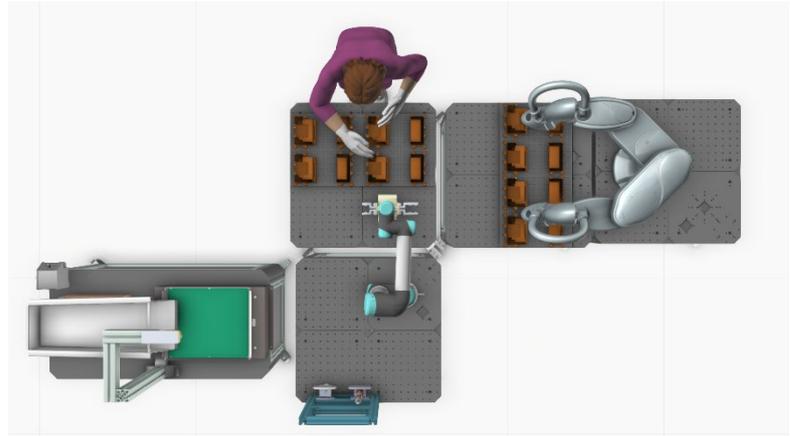

*Figure 14: Modules of production system.*

Modular approach is an ideal solution for reconfigurability especially when combining automation (e.g. robots) and human skills. Off the shelf robots with varying properties of mechanical architecture, degrees of freedom, price, payload and safety limitations may be used to accomplish different types of tasks. Similarly, other hardware if having a universal and modular design can be added or removed when needed. According to the assembly complexity and work tasks distribution between human and robot, modules can be added forming an assembly cell.

### 6.3.2. Standardized and rapid to deploy hardware

General purpose and ready to deploy automation solutions can make it convenient to upscale automation level and boost production volume. It is an ideation of having production system or sub-systems that are portable and quick to implement. Factory in a box concept characterizes standardized modular production units that are flexible, quick to deploy and mobile [68]. The concept of standardized and ready to deploy hardware is available in military combats and health service units but is hard to find in manufacturing landscape. Some hardware solutions are getting available with ease of mobility, and reconfigurability (Figure 15). Such hardware may be utilized for developing reconfigurable production systems.

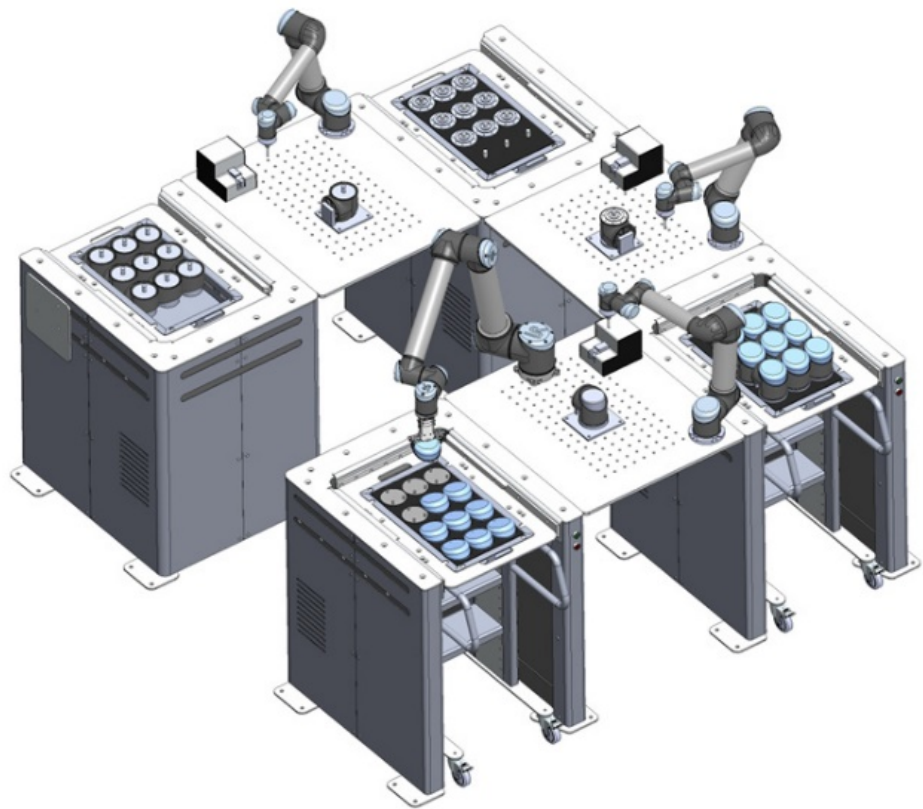

*Figure 15: Standard and modular hardware* [69].

### 6.3.3. Software templates

The most common tasks for robot automation are palletizing, pick and place, machine tending, screwing etc. The robots are programmed in a machine language to perform any of the assigned tasks. However, a large part of the robot program can be used as a template to reprogram the robot if encountered changes e.g. in product design or layout (change in pick and place locations). Ready to use standard program templates can support the end users for faster integration and reconfiguration. An example of this is *EasyTemplates* developed by Danish Technological Institute (DTI) [70] for UR robots. These are easy to use pre-developed templates for most common robotic tasks (Figure 16).

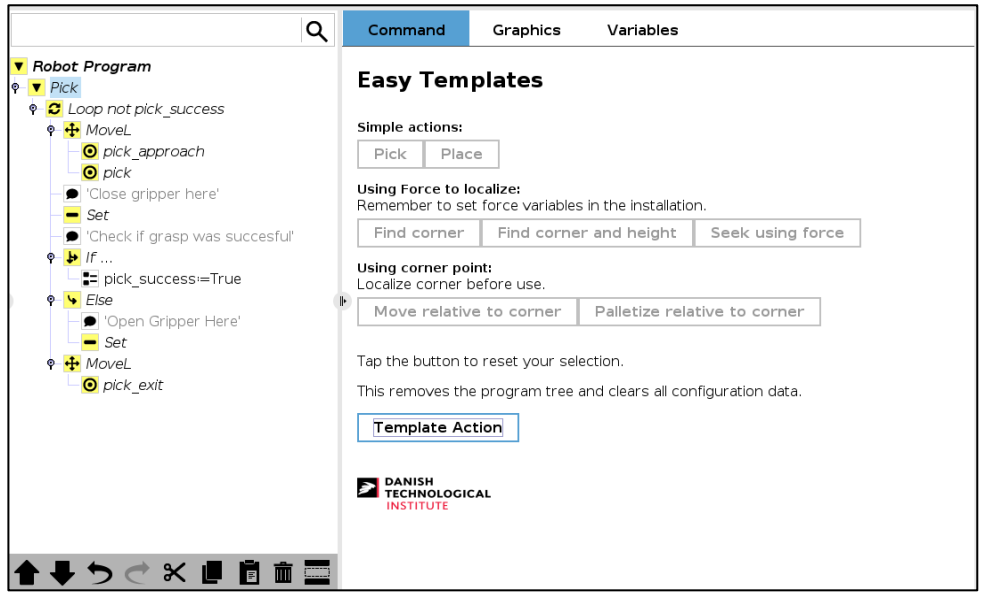

*Figure 16: Templates for cobot program development developed by Danish Technological Institute.*

### 6.3.4. Digital twins

Digital twin (DT) is a data connected representation of design and elements of a system. In the presented work, a DT can be an extension of the simulation models beyond the design phase of the HRC system (Figure 17). DT can support decision making in design, planning and operation of manufacturing systems [6]. Also, the usefulness of digital twins to address several aspects of cobots has been discussed in the literature [4,7]. Simulation-based DTs offer a safe space to get insights into the operational behavior of a complex system, making optimizations and validation of the solutions before bringing solutions into the real world [8].

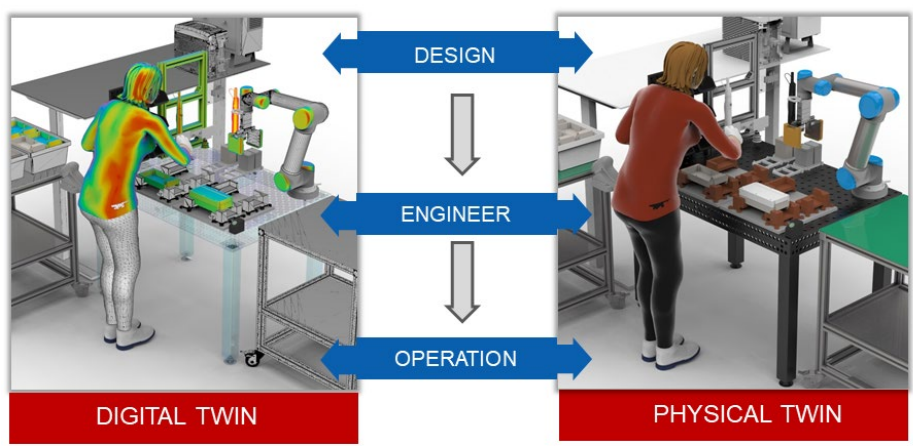

*Figure 17: Digital twin for an HRC assembly system.*

## 7. CONCLUSION

The scale and the severity of the pandemic call the scientific community to guide national and international efforts to combat the disease. Manufacturing sector is under pressure and is struggling to effectively respond to the momentous but hopefully short-term need of products, e.g. a medical ventilator. The flexibility of most manufacturing systems seen today to adapt to product variations is constrained within a very limited boundary. The results of this article prove that cobots with combination of modern manufacturing technologies can be useful for production ramp up in emergency situations.

In this article, the case of using cobots for ventilator production is presented. The study is simulated and validated with a combination of continuous simulation and discrete event simulation. It is validated that collaborative robots can help to form reconfigurable manufacturing systems and can be exploited for ramping up of production when the world faces huge challenge of meeting the unprecedented demand. Different technologies such as modularization, ready to deploy hardware, software templates and digital twins can be combined for faster integration, reconfiguration, and safety validation.

As in the time of emergency demand hike for certain products, manufacturers are under pressure to respond quickly while delivering uncommon results, the measures adapted are often unconventional. As a result, the adapted techniques set new directions for the manufacturing sector. The unconventional approaches may provide new understandings and quite often become a new normal of producing things. The learnings from using cobots during a pandemic, to replace human efforts, may potentially be a step forward towards human-robot teams in factories of the future.